\documentclass[sigconf]{acmart}

\AtBeginDocument{%
\providecommand\BibTeX{{%
Bib\TeX}}}

\usepackage{mathtools} 
\usepackage{bbm}
\usepackage{amsmath}
\usepackage{utfsym}

\usepackage{balance} 

\AtBeginDocument{%
  \providecommand\BibTeX{{%
    \normalfont B\kern-0.5em{\scshape i\kern-0.25em b}\kern-0.8em\TeX}}}

\copyrightyear{2024}
\acmYear{2024}
\setcopyright{rightsretained}
\acmConference[SIGIR '24]{Proceedings of the 47th International ACM SIGIR Conference on Research and Development in Information Retrieval}{July 14--18, 2024}{Washington, DC, USA}
\acmBooktitle{Proceedings of the 47th International ACM SIGIR Conference on Research and Development in Information Retrieval (SIGIR '24), July 14--18, 2024, Washington, DC, USA}
\acmDOI{10.1145/3626772.3657911}
\acmISBN{979-8-4007-0431-4/24/07}
\makeatletter
\gdef\@copyrightpermission{
\begin{minipage}{0.3\columnwidth}
\href{https://creativecommons.org/licenses/by/4.0/}{\includegraphics[width=0.90\textwidth]{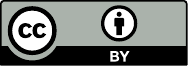}}
\end{minipage}\hfill
\begin{minipage}{0.7\columnwidth}
\href{https://creativecommons.org/licenses/by/4.0/}{This work is licensed under a Creative Commons Attribution International 4.0 License.}
\end{minipage}
\vspace{5pt}
}
\makeatother

\acmSubmissionID{1932}

\begin{document}


\title{RLStop: A Reinforcement Learning Stopping Method for TAR}

\author{Reem Bin-Hezam$^{1,2}$}
\email{rybinhezam@pnu.edu.sa}
\orcid{0000-0002-9156-6186}
\affiliation{%
  \institution{$^1$Department of Information Systems\\College of Computer and Information Sciences\\Princess Nourah Bint Abdulrahman University}
  \city{Riyadh}
  \country{Saudi Arabia}
}

\author{Mark Stevenson$^2$}
\email{mark.stevenson@sheffield.ac.uk}
\orcid{0000-0002-9483-6006}
\affiliation{%
  \institution{$^2$Department of Computer Science\\ Faculty of Engineering\\ University of Sheffield}
  \city{Sheffield}
  \country{United Kingdom}}

\renewcommand{\shortauthors}{Reem Bin-Hezam and Mark Stevenson}

\begin{abstract}

We present RLStop, a novel Technology Assisted Review (TAR) stopping rule based on reinforcement learning that helps minimise the number of documents that need to be manually reviewed within TAR applications. 
RLStop is trained on example rankings using a reward function to identify the optimal point to stop examining documents. 
Experiments at a range of target recall levels on multiple benchmark datasets (CLEF e-Health, TREC Total Recall, and Reuters RCV1) demonstrated that RLStop substantially reduces the workload required to screen a document collection for relevance. RLStop outperforms a wide range of alternative approaches, achieving performance close to the maximum possible for the task under some circumstances.

\end{abstract}



\maketitle

\section{Introduction}

Identifying all, or a significant proportion of, the relevant documents in a collection has applications in multiple areas including identification of scientific studies for inclusion in systematic reviews \cite{higgins2019cochrane,kanoulas2017clef,kanoulas2018clef,kanoulas2019clef}, satisfying legal disclosure requirements \cite{mcdonald2020accuracy,baron2020providing,grossman2016trec}, social media content moderation \cite{yang2021tar} 
and test collection development \cite{Losada2019}. These problems often involve large collections where manually reviewing all documents would be prohibitively time-consuming. Technology Assisted Review (TAR) develops techniques to support these document review processes, including stopping rules which help reviewers to decide when to stop assessing documents, thereby reducing the effort required to screen a collection for relevance. 

TAR stopping rules aim to identify when a desired level of recall (the {\it target recall}) has been reached during document review, while also minimising the number of documents examined. The problem is challenging since these two objectives are in opposition; increasing the number of documents examined provides more information about whether the target has been reached.

A common approach involves estimating the total number of relevant documents in the collection and therefore whether the target recall has been reached, e.g. \cite{shemilt2014pinpointing,Howard2020,callaghan2020statistical,cormack2016scalability,yu2019fast2,Yang2021Heuristic,molinari2023saltau}. Alternative approaches include randomly sampling documents until a sufficient number of relevant ones have been discovered to guarantee that the target recall has been reached \cite{cormack2016engineering,yang2021minimizing} and observing the rate at which relevant documents occur within a ranking until it drops below a pre-defined threshold \cite{cormack2016engineering}. 

Existing methods suffer from a number of limitations. Several approaches carry out repeated statistical testing to determine whether the target recall has been achieved, e.g. \cite{li2020stop,shemilt2014pinpointing,Howard2020,callaghan2020statistical,Yang2021Heuristic}, but this type of sequential testing is statistically invalid \cite{yang2021minimizing}. \citet{lewis2021certifying} avoided this problem using techniques from manufacturing quality control \cite{grant1980statistical} but their approach often required more documents to be reviewed than alternative methods \cite{stevenson2023stopping}. Some methods, e.g. \cite{cormack2016engineering,yang2021minimizing}, fail to take account of the fact that standard TAR workflows \cite{cormack2014evaluation,cormack2016scalability,li2020stop} are highly effective at prioritising relevant documents, meaning that their distribution within the ranking is non-uniform. Exploiting this fact reduces the number of documents that need to be examined before a stopping decision is made \cite{li2020stop,stevenson2023stopping,cormack2016scalability}. 
However, existing approaches to modelling the distribution of relevant documents rely on a particular rate function (e.g. power law \cite{zobel1998reliable} or AP Prior \cite{aslam2005measure}) but the choice of rate function is a modelling decision and may not be appropriate in all circumstances.  

Approaches to TAR stopping essentially involve repeated decisions to either stop or examine more documents. Reinforcement learning (RL) is designed for such sequential decision-making scenarios and has been widely applied within Information Retrieval, e.g. \cite{zeng2018multi,zhou2020rlirank,rosset2018optimizing,nogueira2017task,montazeralghaem2020reinforcement}. However it has not previously been applied to TAR stopping. 
This paper proposes a novel TAR stopping method based on RL. This approach does not rely on invalid statistical assumptions and is able to model rankings to make informed decisions about when to stop. 
Experiments using a range of target recall levels on multiple benchmark datasets demonstrate that the proposed method is able to identify suitable stopping points and performs well in comparison with several previously reported approaches. 

The contributions of this paper are: (1) introduces a novel approach to the TAR stopping problem that makes use of RL, (2) evaluates the proposed algorithm using a range of collections commonly used for TAR problems, and (3) demonstrates that the proposed approach effectively identifies an appropriate stopping point and outperforms a wide range of alternative methods, often substantially.\footnote{Code for the experiments available from \url{https://github.com/ReemBinHezam/RLStop/}}


\section{Approach}

RL is a decision-making method in which an agent aims to maximise the reward obtained from interacting with an environment and making sequential decisions through trial-and-search \cite{sutton2018reinforcement}. The agent's actions are guided by a learned policy ($\pi$) which maps states in the environment to actions. 

RL is applied to the TAR stopping problem by considering an agent that examines a ranked list of documents with the aim of stopping when a predefined target recall has been achieved. The agent examines the ranking sequentially, starting with the highest ranked documents and working down the ranking. For efficiency, documents are examined in batches with relevance judgements obtained for the entire batch simultaneously. After a batch of documents has been examined the agent can either stop examining documents (if it judges that the target recall has been achieved) or continue to the next batch of documents (if not). This approach is similar to previous approaches of the stopping problem in which a ranked list of documents is examined sequentially (e.g. \cite{hollmann2017ranking,di2018study, stevenson2023stopping}), although the stopping decision is made by the agent (following a policy) rather than according to some alternative criteria.

\subsection{RLStop}

This section describes how RL is applied to the stopping problem by outlining how the key elements of an RL system are implemented within it. 

\noindent{\bf State Space:} A ranking is split into $B$ fixed size batches containing $\frac{N}{B}$ documents for a collection of $N$ documents. The agent examines batches sequentially and obtains relevance judgements for all documents in the batch simultaneously. The initial state for each ranking, $S_{1}$, occurs when the first batch (but none of the subsequent batches) has been explored. Additional batches are examined in subsequent states, i.e. in the $n$th state, $S_{n}$, the first $n$ batches have been examined. The final state, $S_{B}$, represents the situation in which the entire ranking has been examined. If the agent reaches this state, it will always stop here since all documents in the ranking have been examined.

States are represented by a fixed size vector of length $B$ in which each element represents a batch. For batches that have been examined the corresponding element shows the number of relevant documents within the batch, while elements corresponding to unexamined batches are given a dummy value (-1). 

\noindent{\bf Action Space:} At each point in the ranking, the agent has a choice between two discrete actions: STOP and CONTINUE. The first action is chosen when the agent (informed by the policy) judges that the target recall has been reached. The stopping point returned is the end of the last batch that has been examined so far. If the agent does not stop it continues to examine the ranking, i.e. moves from state $S_{i}$ to $S_{i+1}$.

\noindent{\bf Reward function:} A reward function, $R(S_{i})$, assigns a score to $S_{i}$ indicating its attractiveness for the agent.  
The reward function is used to train the policy and designing a suitable one is therefore important. A suitable function should: 1) encourage the agent to continue examining documents until the target recall has been reached, 2) discourage further examination after it has been reached and 3) be independent from each topic's specific properties (e.g. total number of documents, ranking shape). In addition, continuous functions are more straightforward for RL algorithms to optimise.

The following function achieves these goals: 

\begin{equation}
R(S_{i}) = 
\begin{dcases*}
   1 - \frac{i}{T} & if $i \leq T$\\   
   - \frac{i - T}{B  - T} & if $i > T$
\end{dcases*}
\end{equation}

where (as above) $B$ is the number of batches into which the ranking is split, $i$ is the index of the current state (i.e. $i$th batch) and $T$ is the batch at which the target recall is reached. (Note that while the value of $T$ is known while the RL algorithm is being trained, it is not known when it is applied.) The function assigns a positive reward when the current state is at, or below, the target recall and a negative reward when it has been exceeded.

The cumulative reward for an RL episode (i.e. examining a ranking until a stopping decision is made) is the sum of rewards for all positions examined by the agent from $S_{1}$ to the stopping step $S_{s}$, i.e. $\sum_{i=1}^{s} R(S_{i})$

\noindent{\bf Policy:} RL aims to learn a policy, $\pi(s, a)$, that maximises the expected cumulative reward obtained by taking an action ($a$) given a state ($s$). Since the state space for our problem is high-dimensional, a neural network is a choice for the policy. The policy is a feed-forward network consisting of an input layer of length $B$, representing the current state, two 64-node hidden layers and a binary output layer indicating the chosen action, which is converted to a probability distribution over actions by a softmax activation function.

\noindent{\bf RL Algorithm:} RLStop uses  Proximal Policy Optimization (PPO) \cite{schulman2017proximal}, a policy gradient approach to RL. PPO is an actor-critic RL algorithm that combines policy-based (actor) and value-based (critic) RL, where the actor decides the actions, and the critic evaluates them. It is based on the REINFORCE \cite{williams1992simple} algorithm but with several enhancements, including the employment of parallel actors running independently by collecting trajectories of different environments simultaneously which allows a policy to be trained using multiple rankings. PPO is also more sample-efficient than some alternative methods, such as DQN \cite{mnih2013playing}, thereby reducing the amount of data required to learn effective policies.

\noindent{\bf Implementation:} The Stable-Baseline3 library \cite{stable-baselines3} was used to implement RLStop. The RL environment was created using the Gymnasium library \cite{gymnasium_2023} which allows multiple environments to be stacked, thereby allowing simultaneous training on multiple topics to ensure the agent is as general as possible. The number of batches, $B$, was set to 100. 


\section{Experiments}

Experiments were carried out on multiple datasets and evaluation metrics. 

\subsection{Datasets}
Performance was evaluated on six datasets widely used in previous TAR work and representing multiple domains. All datasets are highly imbalanced, with a very low percentage of relevant documents per topic. \\
\noindent{\bf CLEF 2017/2018/2019} \cite{kanoulas2017clef, kanoulas2018clef,kanoulas2019clef}: Collections of systematic reviews produced for the Conference and Labs of the Evaluation Forum (CLEF) 2017, 2018, and 2019 e-Health lab Task 2: Technology-Assisted Reviews in Empirical Medicine. The CLEF 2017 dataset contains 42 reviews; CLEF 2018 contains 30 and CLEF 2019 contains 31. RLStop was trained using the 12 reviews provided with the CLEF 2017 dataset.\\
\noindent{\bf TREC Total Recall (TR)} \cite{grossman2016trec} A collection of 290,099 emails associated with Jeb Bush’s eight-year tenure as Governor of Florida (athome4). The collection contains 34 topics. RLStop was trained using the athome1 dataset consisting of 10 topics.\\
\noindent{\bf RCV1}  \cite{lewis2004rcv1} A collection of Reuters news articles labelled with subject categories.  Following \cite{yang2021minimizing, Yang2021Heuristic}, 45 categories were used to represent a range of topics, and the collections were downsampled to 20\%. RLStop was trained using the remaining RCV1 topics, excluding those already included in the test set.

Each collection was ranked with the use of a reference implementation \cite{li2020stop} of AutoTAR \cite{Cormack2015}, a greedy Active Learning approach representing state-of-the-art performance on total recall tasks widely used for TAR experiments. The use of AutoTAR rankings allows direct comparison between RLStop and alternative approaches used as baselines in previous work \cite{li2020stop,stevenson2023stopping}. 

\subsection{Evaluation Measures}

A wide range of metrics have been used to evaluate TAR stopping methods, e.g. \cite{Yang2021Heuristic,li2020stop}. These essentially measure the method's success of meeting the two objectives of stopping algorithms: (1) reach the target recall and (2) examine as few documents as possible. 
Two metrics were used to capture these objectives. {\bf Recall}: the proportion of relevant documents identified by the method.
{\bf Cost}: percentage of documents examined. Results for these metrics are reported as average scores for all topics in each collection. 

An additional metric captures elements of both objectives and is used to quantify the variation in performance across topics contained within a collection. {\bf Excess}: proportion of documents examined after the target recall has been reached or that needs to be examined reach it \cite{binhezam_stevenson_emnlp23}. Excess is defined as follows: 
\begin{equation}\label{eq:e_cost}
excess = \frac{cost(method) - cost(optimal)}{1-cost(optimal)}
\end{equation}
where $cost(method)$ and $cost(optimal)$ are the cost of the method being evaluated and stopping at the optimal point (i.e. oracle). An excess of 0 indicates that the optimal stopping point for the target recall has been reached, a positive score indicates that more documents than absolutely necessary were examined (i.e. the target was overshot) and a negative indicates that the target was not reached (i.e. undershot). 

Metrics were calculated using  the {\tt tar\_eval} open-source evaluation script.\footnote{https://github.com/CLEF-TAR/tar} 

\subsection{Training and Hyper-parameter Tuning}

RLStop models were trained for 100,000 timesteps on the each target recall using each dataset's training split. PPO hyper-parameters were set via a grid search on the CLEF 2017 training dataset and the following values chosen: 
batch size = 100, number of steps (used to collect trajectories before each policy update rollout) = 100, learning rate = 0.0001, number of epochs = 8, entropy coefficient (which encourages policy exploration) = 0.1, discount factor = 0.99 and KL coefficient (which controls the clipping range) = 0.2.

\subsection{Baselines and Oracle}

RLStop was compared against a range of alternative approaches used as baselines in previous work \cite{li2020stop,stevenson2023stopping}. The target method (TM) \cite{Cormack2016} randomly samples documents until a set of predefined target set is identified with results reported for the original method \cite{Cormack2016} and two extensions: TM-adapted \cite{stevenson2023stopping} and QBCB \cite{lewis2021certifying}. SCAL \cite{cormack2016scalability} and AutoStop \cite{li2020stop} estimate the number of relevant documents by sampling across the entire ranking. SD-training/SD-sampling 
  \cite{hollmann2017ranking} are score distribution methods. The knee method \cite{cormack2016engineering} identifies the inflection point in the gain curve. IP-H \cite{stevenson2023stopping} uses a counting process to estimate the number of relevant documents. 

Baselines are computed using reference implementations from previous work \cite{li2020stop, stevenson2023stopping} where possible. Otherwise, previously reported results are used and are directly comparable since they are also based on AutoTAR rankings. 
However, some baselines are not available for RCV1 since we were unable to run the reference code and results have not been provided in previous work. 

Performance was also compared against an Oracle method (OR) which examines documents in ranking order and stops when the target recall level has been achieved (or exceeded). The oracle represents the behaviour of an ideal stopping method but is not useful in practise since it requires full information about the ranking.\footnote{Note that the recall achieved by the oracle is higher than the target recall when it is not possible to achieve the target recall exactly, e.g. given a target recall of 0.8 and collection containing 9 relevant documents, the oracle will stop after 8 relevant documents have been found, i.e. recall 0.89.}

\section{Results and Analysis}

\begin{figure*}[!htb] 
  \centering
  \includegraphics[width=0.974\linewidth]{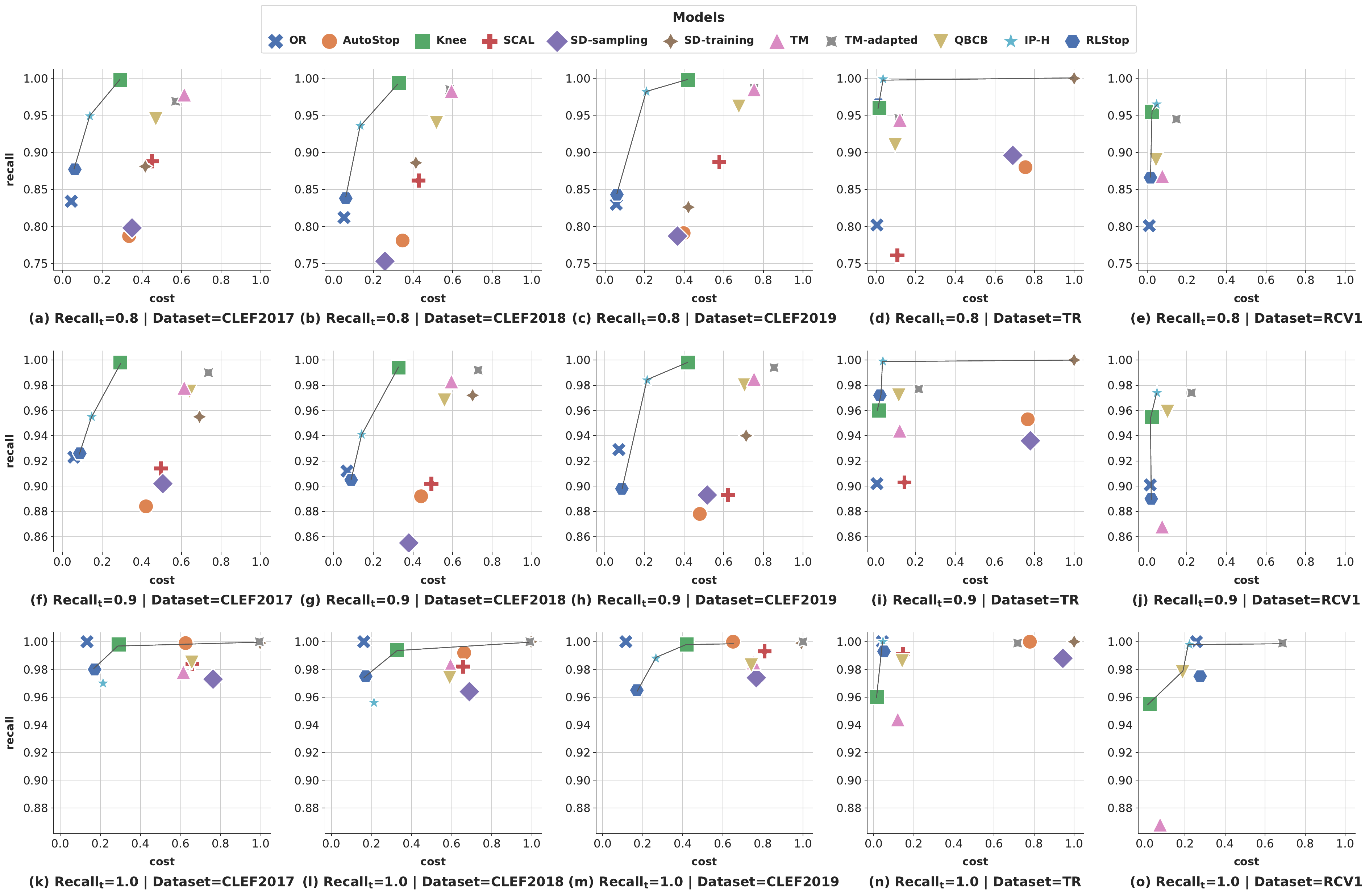} \caption{Performance of RLStop and baselines for Recall vs. Cost metrics. Grey lines indicates non-oracle Pareto optimal approaches. (Note differences in range of y-axis (recall) to avoid clustering of results.)} 
  \label{fig:BLs_compare_recall_cost}
  \Description{RLStop vs baselines}
\end{figure*}

Figure~\ref{fig:BLs_compare_recall_cost} shows the recall and cost for RLStop and alternative methods for each data set at multiple target recall levels: \{1.0, 0.9, 0.8\}. 
Performance of RLStop (denoted by blue hexagon) is often close to the optimal oracle results (e.g. sub-figures (b), (c), (f), (g), (j) and (n)) and is also Pareto optimal\footnote{No other approach achieves the same, or greater, recall with lower cost.} 
in the majority of cases.
RLStop is not Parteo optimal in two cases (sub-figures (n) and (o)) but is very close to the Pareto frontier both times. Other baselines that are commonly Pareto optimal include IP-H and the Knee method (cyan star and a green square, respectively). However, both methods tend to overshoot the target more than RLStop. Other baselines are substantially more costly and either undershoot or overshoot the target more frequently than RLStop.

Although RLStop tends to follow the target recall it does not do so exactly and tends to overshoot for lower target recalls (e.g. 0.8). This overshooting is particularly pronounced for the TR dataset where the target recall is often reached very quickly (as demonstrated by the oracle performance). For some topics overshooting is caused by the ranking being divided into fixed batches (1\% of the ranking) where examining only a single batch (i.e. the earliest possible stopping point) overshoots the target recall. Increasing the number of batches may be a potential solution to this problem. 
On the other hand, RLStop tends to undershoot for higher target recall levels, although it normally stops within a few percentage points of the target. This undershooting may be due to topics that contain a small number of relevant documents towards the end of their rankings, which RLStop does not reach.

\begin{figure}[!htb] 
  \centering
  \includegraphics[width=1.0\linewidth,trim=0 5 0 40, clip]{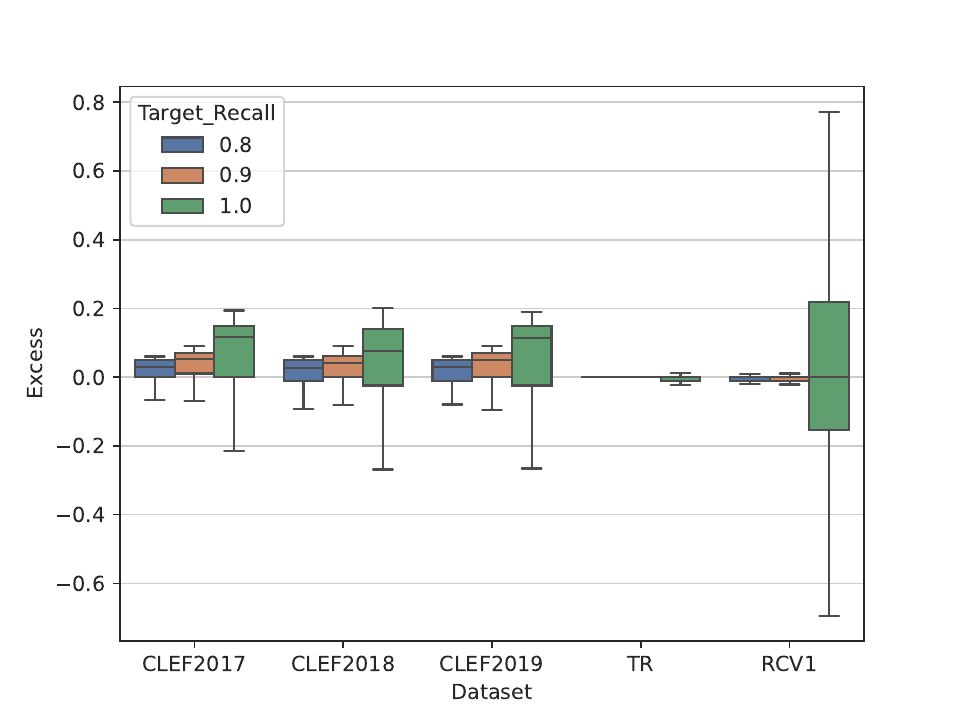}  
  \caption{Distribution of RLStop excess across topics. (Outliers for target recall 1.0 removed for clarity.)} 
    \Description{RLStop excess across topics}
  \label{fig:excess_all}
\end{figure}

Figure \ref{fig:excess_all} shows how RLStop's excess varies across topics for all datasets at different target recall levels. 
For the majority of topics the excess costs is confined within a fairly narrow range, particularly for the TR collection. The variation is higher for all collections when target recall is 1.0 due to the additional challenge of identifying all relevant documents. This is more pronounced for RCV1, although the excess is within a relatively narrow range for the majority of topics.
 
\section{Conclusion}

This paper proposes RLStop, a novel TAR stopping rule based on reinforcement learning. RLStop substantially reduces the workload required to screen collections for relevance. RLStop performs well in comparison with several baselines on multiple benchmark datasets at different target recall levels.

RLStop requires training data which may be available (e.g. from previous relevance screening carried out within a similar environment, such as within a systematic review team). It may not be suitable if these are not available or if confidence guarantees of reaching the target recall are required. RLStop also requires a new model to be trained for each target recall level. We plan to address these issues in future work. 


\clearpage

\bibliographystyle{ACM-Reference-Format}
\balance
\bibliography{references}

\end{document}